\documentclass[fleqn,10pt]{wlscirep}
\usepackage[utf8]{inputenc}
 \usepackage[T1]{fontenc}
\usepackage{multirow}
\usepackage{array}
\usepackage{booktabs}
\title{Exploring the effectiveness of a COVID-19 contact tracing app using an agent-based model}

\author[1]{Jonatan Almagor}
\author[1,*]{Stefano Picascia}
\affil{MRC/CSO University of Glasgow, Social and Public Health Science Unit, Berkeley Square, 99 Berkeley Street, Glasgow, G3 7HR, Scotland}
\affil[*]{stefano.picascia@glasgow.ac.uk}

\begin{abstract}
A contact-tracing strategy has been deemed necessary to contain the spread of COVID-19 following the relaxation of lockdown measures. Using an agent-based model, we explore one of the technology-based strategies proposed, a contact-tracing smartphone app. The model simulates the spread of COVID-19 in a population of agents on an urban scale. Agents are heterogeneous in their characteristics and are linked in a multi-layered network representing the social structure - including households, friendships, employment and schools.
We explore the interplay of various adoption rates of the contact-tracing app, different levels of testing capacity, and behavioural factors to assess the impact on the epidemic. 

Results suggest that a contact tracing app can contribute substantially to reducing infection rates in the population when accompanied by a sufficient testing capacity or when the testing policy prioritises symptomatic cases. As user rate increases, prevalence of infection decreases. With that, when symptomatic cases are not prioritised for testing, a high rate of app users can generate an extensive increase in the demand for testing, which, if not met with adequate supply, may render the app counterproductive. This points to the crucial role of an efficient testing policy and the necessity to upscale testing capacity. 
\end{abstract}
\begin{document}

\flushbottom
\maketitle

\thispagestyle{empty}

\section*{Introduction}

At the time of writing Coronavirus Disease 2019 (COVID-19) has caused
just over 1,200,000 confirmed deaths in 216 countries worldwide \cite{WHOData2020}.
In the absence of an effective treatment, let alone a vaccine, the
only possible mitigation strategies are non-pharmaceutical \cite{Ferguson2020}.
In the initial phase of the pandemic lockdown measures, implemented
with various degrees of firmness in most countries, have proven effective
in containing the epidemic \cite{Flaxman2020a} and reducing the basic
reproduction number $R_{0}$ to less than 1. However, these measures
cannot be sustained for a prolonged period of time, as the economic
damage inflicted on workers, enterprises and governments would be
irreparable. Therefore, a second phase of containment has followed,
in which lockdown measures are lifted and different mitigation strategies
are required, namely case isolation, tracking and contact-tracing,
which are rooted in the established epidemiological toolkit \cite{Fraser2004}.
These are particularly crucial in the context of a disease that is
transmitted by a large proportion of a-symptomatic and pre-symptomatic
individuals, as studies of early outbreaks in China and Italy have
indicated \cite{Lavezzo2020,Gandhi2020,Bai2020,Tong2020}. However,
manual contact tracing can be a time-consuming and inefficient exercise,
since models show that the probability of epidemic control decreases
rapidly when not enough cases are ascertained via contact-tracing
before the onset of symptoms \cite{Hellewell2020}. Technology-based
solutions have been proposed to automatise the tracking process, in
the form of contact tracing smartphone apps. Exploiting the Bluetooth
Low Energy technology, these applications can trace other smartphones
coming into close contact for a period of time compatible with potential
infection transmission \cite{Zastrow2020}. Once an individual who
uses the app discovers having been infected, all smartphones who have
come into close contact receive a notification, signalling the potential
exposure. Intense debate followed in the scientific community and
among the public on the risks, especially to privacy, and effectiveness
of the app-based solution \cite{Sweeney2020}. A number of studies,
using a range of methodologies, tried to establish the optimal adoption
rate and the other necessary measures (such as social distancing and
testing) required for the app to be effective in containing or suppressing
the epidemic \cite{Braithwaite2020Review}. Analytical mathematical models show a generally optimistic stance, suggesting that instantaneous contact tracing, such as that afforded by the app-based solution, can lead to epidemic control \cite{Ferretti2020a},
if at least 60\% adoption rate of the app is attained \cite{Fraser2020}. 

Alternatively, other models highlight possible unintended consequences of the mitigation strategies. For example, adopting the app with an inadequate number of tests available may lead to epidemic control, but at the price of an unrealistic number of people having to isolate \cite{Firth2020}. Another model \cite{SIMASSOC2020},
still in the development phase, seems to suggest that the app might
not be effective at all, and other measures could be preferable, such
as a random testing policy. 

Here we present an agent-based model designed to explore the effectiveness
of the app in containing or suppressing the epidemic. The model simulates
the complex interplay between: (1) the portion of the population that
uses the app, (2) the availability of testing, and (3) behavioural
factors, such as the willingness to comply with self-isolation instructions.
The model explicitly simulates realistic interactions and examines
the different policy options on the table. Our modelling approach
tries to balance the competing needs for flexibility and scalability,
while retaining as much complexity and descriptiveness as possible.

\section*{\label{sec:A-model-of} Methods: an agent-based model of COVID-19 transmission and mitigations}

Modelling the effectiveness of mitigation strategies requires that
we simulate them in parallel with the spread of the disease itself.
Building on the principles of the SEIR approach, an agent-based model
(ABM) was developed, simulating the spread of COVID-19 within the population of an urban
area. Since the virus is transmitted through contacts between infected
and susceptible individuals, in order to understand the dynamics of
the spread it is essential to represent the multiple social networks
that connect individuals within a population and determine patterns
of contacts. A major weakness of traditional compartmental infection
models is their aggregated nature. These models divide the population
into homogeneous groups (compartments) in accordance with the state
of the disease (SEIR), and assume disease transmission
to occur as the infected group mixes with the susceptible group at
certain rates \cite{Chitnis2013}. These models' assumptions do not
account for the heterogeneity that exists between individuals within
the groups, and simplify the complexity of contact patterns, which take place in social networks that are important to understanding the course of an epidemic \cite{Nielsen2020}. 

Our ABM intends to bridge this gap by modelling more realistic contact patterns that take place among heterogeneous agents interacting within a social network. Central to the agent-based approach is that each agent is represented in the model individually,
with their specific characteristics (such as age and sex) and behaviour.
Furthermore, interactions between agents are explicitly simulated.

\subsection*{Code availability}
The full source code and datasets employed are open source and maintained at the following address: \url{http://github.com/harrykipper/covid}

\subsection*{Agents' social networks and daily contacts }

We generate a synthetic population of circa 103,000 agents, derived
from the 2011 UK Census \cite{Census2011}, including household type,
gender and age within geographical zones (Detailed Characteristic
Sector 2011). The population represents the city of Glasgow, Scotland.
A multi-layered social network links agents within the following social
structures:
\begin{itemize}
	\item A \emph{household} structure is created as follows: individuals who
	belong to households classed with the same type in the Census, and
	reside in the same locale, are linked together on the basis of age
	difference; single people below the age of 20 are assumed to live
	at home with one or two parents and siblings. Single people above
	the age of 20 are assumed to live independently, with a certain proportion
	co-habiting. Links of type \textquoteleft household\textquoteright{}
	are built among these agents.
	\item Family \emph{relatives} who don't live in the same household (i.e.
	grandparents) are linked together.
	\item Several \emph{workplace} sites are created, based on the distribution
	of workplace sizes in the city of Glasgow \cite{ScotBusiness}. Active
	working-age agents are distributed among workplaces and linked to
	all co-workers at the same site as well as to a subset of colleagues
	who are assumed to be in closer, more frequent contact. Out of the
	working-age population, 13\% of agents are assigned to customer-facing
	employment \cite{ONS2018}, experiencing frequent contact with random
	agents of the population during work.
	\item A \emph{friendship} network links agents over 14 years of age, generated
	following the Barabasi-Albert model \cite{Barabasi1999} so that a
	scale-free network is produced, characterised by variation in number
	of friends per agent, with a median of 14 friends per agent, skewed
	towards similar age.
	\item Children between 6-17 years of age also belong to \emph{classes} of
	maximum 30 children of the same age from the same zone and are linked
	together as classmates. 
\end{itemize}
Using the social network we simulate daily contacts that take place between individuals within a population. Each type of social domain (household, workplace/school, friends, relatives and random contacts) is defined by frequency of encounters and the number of contacts per encounter (Table \ref{tab:Contact-type-and}).    
Agents come into contact with all household members daily, and meet with relatives twice per week. They attend the workplace and school 5 days per week. For school, we simply assume that during a day, agents have contacts with half of their classmates, which are randomly selected. For the workplace, agents have contacts with a group of close colleagues and with one randomly selected co-worker. In this way workplace employees are clustered in small groups having daily contacts (representing colleagues working in proximity) but also interact less frequently with others in the larger workplace. As for social encounters, it is highly difficult to define a typical number of friends. Here, we wish to account for the heterogeneity that exists between highly socially active individuals, having many friendship ties, and less active ones, having only few social ties. Therefore, for each meeting the number of contacts is randomly selected between 1 friend to 10\%\ of the agent’s friendship ties. Furthermore, we assume that elderly agents have a reduced frequency of social encounters \cite{kwok2014social,Mossong2008}. 
As for random contacts with strangers, we assume the number of contacts is a proportion $p$ of the size of the area of residence (zone size range between 200-2700, with a mean of 1100 agents). Hence, each day agents have random contacts with $p$ = 1\%\ ($\sim$11 contacts) of the population residing in their zone, and public-facing workers have contacts with 3$p$= 3\%\ ($\sim$33 contacts). To reflect daily variability in encounters, the number is sampled from a Poisson distribution. 
The emerging distribution of daily contacts in the model resembles the shape of the distribution derived from a UK population contact survey (see: Model calibration and baseline scenario\ref{sec:ModelCalibration}). The impact of different contact frequencies on ABM results is furthered explored in a sensitivity analysis (Supplementary Table S2)    

\begin{table}
	\begin{tabular}{|>{\centering}p{0.15\columnwidth}|>{\centering}m{0.3\columnwidth}|>{\centering}m{0.32\columnwidth}|>{\centering}p{0.15\columnwidth}|}
		\hline 
		Type of contact & Frequency of encounters & No. of contacts per encounter & Transmission probability ($\beta$) per contact\tabularnewline
		\hline 
		\hline 
		Household & Daily & All household members & $\beta_{c}$\tabularnewline
		\hline 
		School & 5 days per week & 50\% of the class & $\beta_{k} = \beta_{c}\times0.5$\tabularnewline
		\hline 
		Friendship / Acquaintance & \multirow{1}{0.3\columnwidth}{Daily (age < 65 years)\foreignlanguage{english}{\linebreak }3.5 days per week (age > 65 years)} & 1-10\% of their friends & $\beta_{c}$\tabularnewline
		\hline 
		Relations & 2 days per week & One relative per household & $\beta_{c}$\tabularnewline
		\hline 
		Workplace & 5 days per week & All close colleagues and one from other colleagues & $\beta_{c}$\tabularnewline
		\hline 
		Workplace (public facing) & 5 days per week & Random contacts are drawn from a Poisson distribution $Pois(\lambda_{W})$
		
		$\lambda_{W}=3$p$\times zone_{-}population$ & $\beta_{r} = \beta_{c}\times0.1$\tabularnewline
		\hline 
		Random & Daily (age < 65 years);\foreignlanguage{english}{\linebreak }3.5
		days per week (age > 65 years) & Random contacts are drawn from a Poisson distribution\foreignlanguage{english}{
			$Pois(\lambda_{R})$\linebreak }
$\lambda_{R}= $p$\times zone_{-}population$ & $\beta_{r} = \beta_{c}\times0.1$\tabularnewline
		\hline 
	\end{tabular}
\caption{Contact type and transmission probability
for social network based and random encounters}
\label{tab:Contact-type-and}
\end{table}

\subsection*{Viral transmission}

During each day, all infected agents come into contact with the subset of agents in each social environment. For each contact with a susceptible agent the virus may be transmitted with a certain probability (Table \ref{tab:Contact-type-and}). The probability of transmission during a contact with an agent in the network ($\beta_{c})$ is higher than during a random contact with a stranger ($\beta_{r}$), as we assume that a contact with a stranger is of shorter duration and reduced closeness, translating in a reduced likelihood of transmission (Table \ref{tab:Contact-type-and}). In accordance with evidence that children are less susceptible to COVID-19 \cite{Davies2020}, we reduce the probability of infection by 50\% for agents under the age of 16. 

\subsection*{Progression of the disease }

\begin{figure}[h]
	\centering
	\includegraphics[scale=0.8]{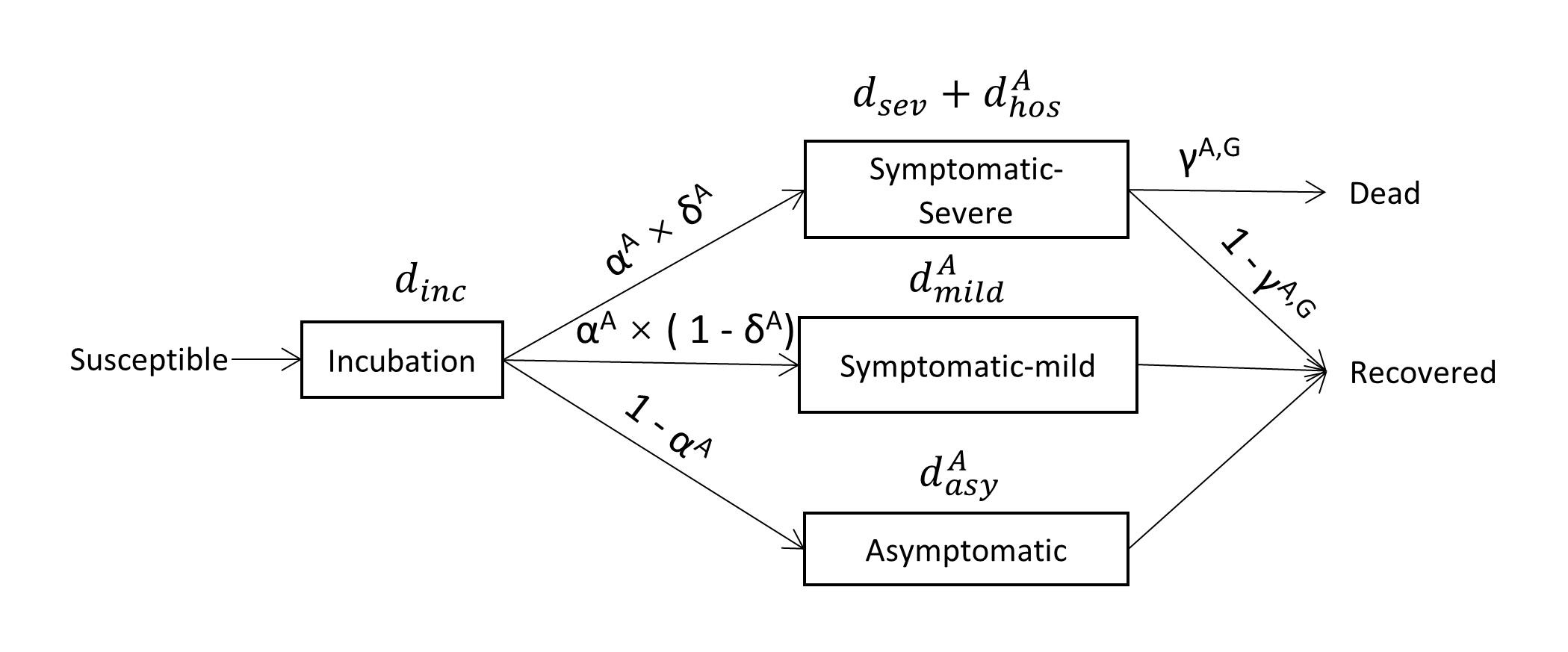}
	
	\caption{Disease progression. Rectangles represent
		states of the disease and arrows the transition between states. $d_{state}^{A}$
		denotes the duration of the disease state given the age $A$ of the
		agent. Agents with severe symptomatic disease spend $d_{sev}$ days
		at home before being admitted to the hospital for a duration of $d_{hos}^{A}$
		days. Transition between disease states occurs with age dependent
		probabilities; where $\alpha^{A}$ and $1-\alpha^{A}$ denote the
		probability of an agent of age $A$ being symptomatic and asymptomatic,
		respectively. $\delta^{A}$ denotes the probability for a symptomatic
		agent to progress into severe disease; $\gamma^{A,G}$ denotes the
		probability of a severely ill agent of age $A$ and gender $G$ to
		die. For details of parameter values see: Supplementary Table S1.}
	\label{fig:Disease-progression}
\end{figure}

Once a susceptible agent is infected, she progresses through the various states of the disease (Figure \ref{fig:Disease-progression}). The progression between disease states and the duration of each state
are based on probabilities and durations which are age and gender dependent, as estimated in recent research on COVID-19 patients (Supplementary Table S1). Initially, the disease is in the incubation phase, a stage in which the agent is not infectious. A fraction of the infected agents becomes infectious 1-3 days before the end of the incubation period (pre-symptomatic infection), while others are infectious only at the
end of incubation \cite{Gandhi2020,Lauer2020,Huang2020}. To reflect
that, during each of the last 3 days of incubation agents have a 25\%
probability of becoming infectious. Following the incubation period,
agents are either asymptomatic or symptomatic. Asymptomatic agents
are able to infect others, but do not feel symptoms, and we assume
that after the 3rd day of being infectious, infectiousness declines
by 10\% each day that follows \cite{Yang2020}. Symptomatic agents
with a mild disease are assumed to feel the symptoms, but do not require hospitalisation; severely ill symptomatic agents initially stay at
home and are then admitted to the hospital. Once severely ill agents
are admitted to the hospital, we assume they do not come into contact
with any other agent. The model does not simulate nosocomial infection
and always assumes the availability of hospital beds. In accordance
with findings, agents stop infecting others after 7-11 days from the
onset of symptoms. \cite{MIN2020}. 

\subsection*{Mitigation strategies: testing, contact tracing app}

The model includes two types of mitigation tools to track and trace infected agents: the contact tracing app (CTA) and COVID-19 detection tests. The CTA is distributed among a fraction of the population aged over 14. It stores in memory the ID of all other CTAs it came into contact with over the course of the previous 10 days. Infected agents who were tested  positive can use the CTA to notify their contacts of possible exposure. 

Symptomatic agents who seek testing are assumed to get tested between 1 and 3 days after the onset of symptoms, and results are determined within a day. We assume that a fixed number of tests is available: as agents are tested the stocks decrease and restocking takes place daily. Agents seek testing when: (a) they feel symptoms, (b) they are notified of possible exposure to an infected agent. A notification is made to exposed agents when: (a) an agent feels symptoms or receives a positive test result and alerts all relatives; (b) a pupil tests positive and all classmates are made aware and quarantined; (c) a CTA user receives a positive test result and the app notifies all recorded contacts (Figure \ref{fig:Procedure-for-testing}). 

To reflect the impact of influenza-like illness (ILI) on the testing system, we assume that over the course of any given week, 3.5\% of the population experience ILI  \cite{FluSurvey2020}, of which 30\% will seek COVID-19 testing. While these agents test negative, they contribute to the depletion of tests.

\subsection*{Agents' adherence to self-isolation instructions}

When agents self-isolate all their social ties are removed, except for household ties, as they are assumed to self-isolate at home. It is likely that some precautions are put in place between the infected agent
and her household members, therefore we assume a 30\% reduction in the probability of transmission within the household. 

\begin{figure}[h]
	\centering
	\includegraphics[scale=0.5]{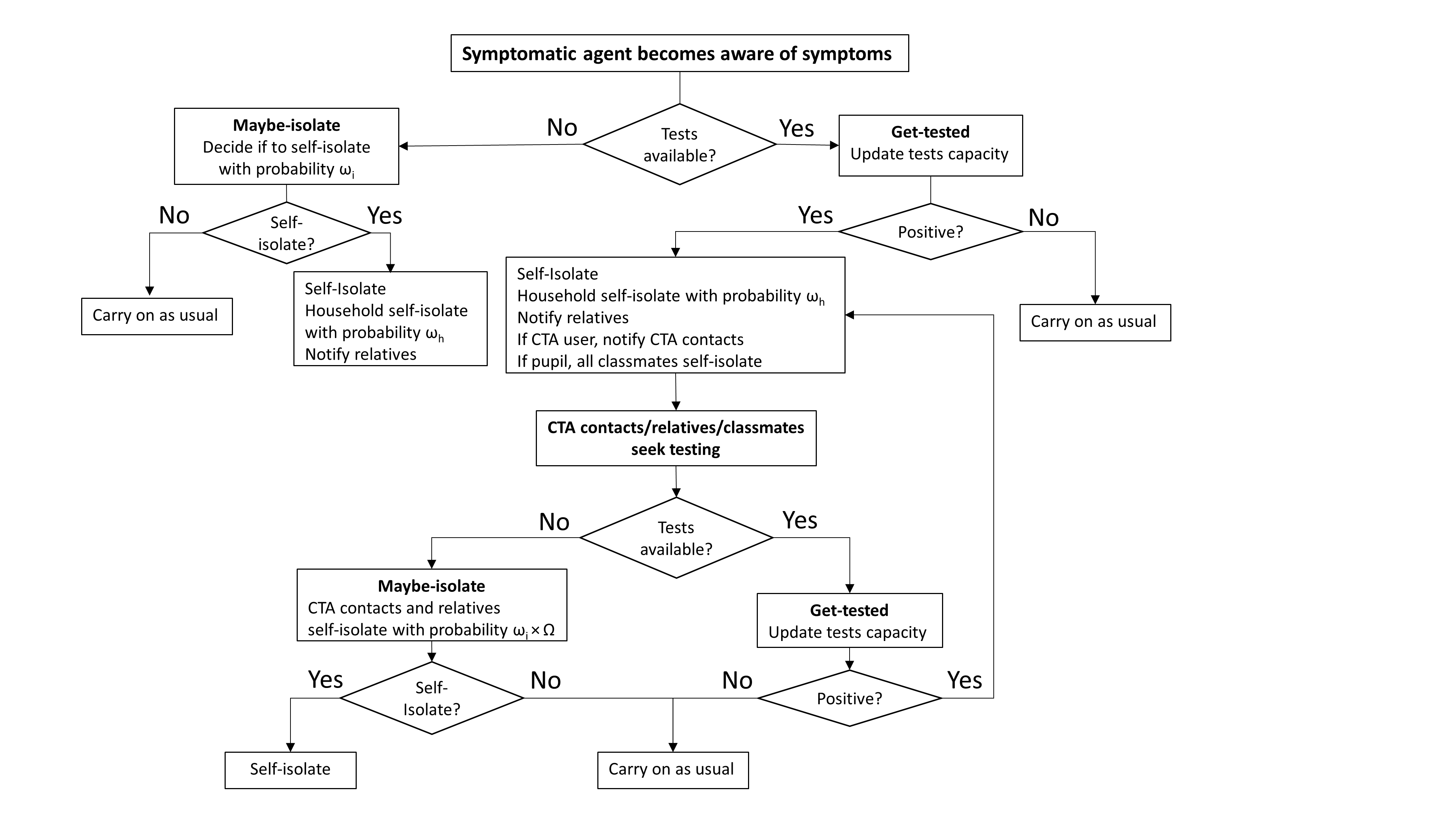}
	\caption{Procedure for testing, self-isolation
and CTA notification. Once symptomatic agent $i$ becomes aware of
the disease, the agent seeks testing. If tests are available, the agent gets tested.
Following a positive result the agent will self-isolate. Household members $h$
of agent $i$ will self-isolate with probability $\omega_{h}$ and relatives will be notified. If the agent is a pupil all classmates will self-isolate and seek testing. If agent $i$ uses the CTA all the recorded contacts  will be notified. When tests are unavailable, agent $i$ will self-isolate with probability $\omega_{i}$. In case the agent self-isolates the aforementioned procedure of self-isolation will take place without the notification to CTA contacts. Otherwise the agent will continue as usual. Once CTA user j is notified, agent j will seek testing. If testing is unavailable, agent j may self-isolate with probability $\omega_{j}*\Omega$ }
	\label{fig:Procedure-for-testing}
\end{figure}

Without the certainty that testing provides, surveys suggest that not all individuals will comply with self-isolation guidelines, both when feeling symptoms or when notified by the CTA \cite{Altmann2020}.
We denote a parameter $\omega_{i}$ representing the probability of agent $i$ to self-isolate when feeling symptoms of COVID-19. We also assume that agents who are notified by the CTA, but do not feel any symptoms, are less likely to self-isolate (than if they had symptoms) without testing. Therefore, their probability to self-isolate is reduced by factor $\Omega$, where $0<\Omega<1$. In the model, the probability of self-isolation varies between agents with mean $\omega=70\%$.
Figure \ref{fig:Procedure-for-testing} presents the algorithm triggered once an agent becomes aware of her symptoms. The procedure triggers a chain of actions performed by symptomatic agents that involves testing (if available), deciding whether to self-isolate and notifying relatives and CTA contacts. Following that, exposed agents who were notified perform similar actions. For a pseudo code of the main model functions see: Supplementary Information. 

\section*{Model calibration and baseline scenario} \label{sec:ModelCalibration}

\begin{figure}
	\includegraphics[scale=0.5]{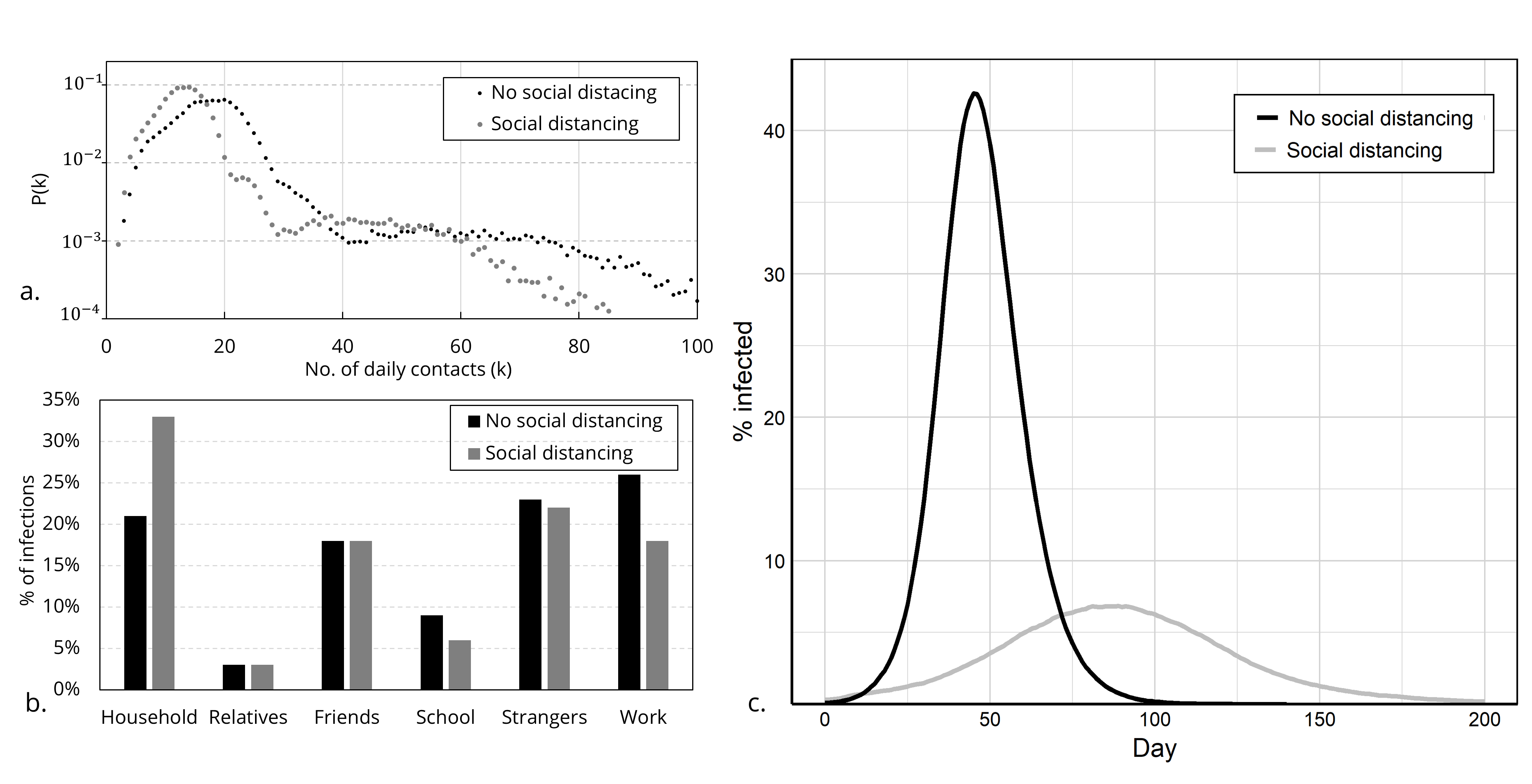}
	
	\caption{Distribution of daily contacts and epidemic dynamics, for scenarios with and without social distancing. (a) Distribution of number of daily contacts. (b) Distribution of infection sources by type of contact. (c) Infection prevalence by day, from beginning to end of the epidemic.}
	\label{fig:Epidemic-dynamics-with}
\end{figure}

The initial scenario reproduces a \textquoteleft business as usual\textquoteright{} situation with no mitigation in place, with contact frequencies as specified in Table \ref{tab:Contact-type-and}. To verify the contact
patterns generated by the ABM, we compared the properties of the distribution of agents\textquoteright{} daily contacts generated by the model to a distribution of contacts derived from a survey conducted in the
UK \cite{Danon2012}. The distribution of daily contacts generated by the ABM holds a similar shape to the distribution in the survey. It is characterised by  a lognormal body where most contacts are between 5-25, and a long tail of higher number of contacts with lower frequencies (Figure \ref{fig:Epidemic-dynamics-with}a). The mean and
standard deviation of number of daily contacts are 20 and 13, respectively.  

To calibrate the model, we tested a range of transmission probability ($\beta_{c}$) values to generate the basic reproduction number of $R_0\sim2.8$ in the initial three weeks of the epidemic, as estimated for the UK \cite{Stedman2020}. The best fit was achieved for $\beta_{c}$=8\%, and $\beta_{r}$=0.8\%. 

After establishing the initial scenario, we simulate the post-lockdown situation expected in several countries, in which most restrictions are lifted but citizens are still encouraged to work from home when possible, limit social interactions, maintain physical distancing and wear face masks in public. Therefore, in this scenario we assume 3 days attendance per week at workplaces and schools; and a reduction of 30\% in contacts in schools, with strangers, as well as the frequency of social meetings (within the \textquoteleft friendship\textquoteright{} network). The contact distribution in this scenario is presented in Figure \ref{fig:Epidemic-dynamics-with}a. The mean number of daily contacts is reduced from 20 to 14. In addition, transmission probability for contacts outside of the household is reduced by 30\% ($\beta_{c}$ = 5.6\% and $\beta_{r}$ = 0.56\% ) to reflect measures such as face mask usage, social distancing and increased hygiene, all of which reduce the likelihood of viral transmission. We refer to this situation as our \emph{baseline} scenario. The reproduction number in this scenario comes down to 1.5. Comparing the scenarios, when social distancing is practised, the proportion of infected agents at the peak of the epidemic is significantly reduced from 43\% to 7\% (Figure \ref{fig:Epidemic-dynamics-with}c). The distribution of the sources of infection also varies, as the proportion of infections originating in workplaces and schools is reduced and the household becomes the predominant locus of transmission (Figure \ref{fig:Epidemic-dynamics-with}b).

\section*{Experimental design}

The core of our study explores the introduction of the CTA and the availability of testing into the baseline scenario of social distancing. We simulate the impact on viral spread of various combinations of: (1) proportion of CTA users in the population; (2) levels of testing capacity; (3) levels of compliance with self-isolation on the part of CTA users; (4) testing policy. Table \ref{tab:Parameters-used-in} summarises the parameter combinations explored in the model. Overall, we simulated 140 scenarios, each repeated 20 times to account for uncertainty in the results due to the stochasticity embedded in the model. 

\begin{table}
	
	\begin{tabular}{|>{\centering}p{0.15\columnwidth}|>{\centering}p{0.4\columnwidth}|>{\centering}p{0.4\columnwidth}|}
		\hline 
		Parameter & Value in experiments & Description\tabularnewline
		\hline 
		\hline 
		CTA users & 0, 20, 40, 60, 80 & \% population over age 14 using CTA\tabularnewline
		\hline 
		Testing capacity & 0, 0.5, 1, 1.5, 3, 6, \emph{unlimited} & Maximum \% of population that can be tested per week (reflective of UK policy announcements of 200,000 tests per day, corresponding to 2\% of the population per week)\tabularnewline
		\hline 
		Compliance of CTA users & Low compliance: $\Omega=0.5$
		
		High compliance: $\Omega=0.9$ & Probability to self-isolate for CTA user without testing decreases
		by $\Omega$: probability to self-isolate = $\omega_{i}*\Omega$\tabularnewline
		\hline 
		Testing policy & a) Priority to symptomatic agents
		
		b) No priority & a) Each day symptomatic agents are tested, only then CTA users
		
		b) Agents are tested on basis of first come first tested\tabularnewline
		\hline 
		Initial conditions & 7\% recovered from the virus; 300 (0.3\%) agents infected & Reflects UK estimates ahead of lifting lockdown\tabularnewline
		\hline 
	\end{tabular}

\caption{Parameters used in simulation experiments}
\label{tab:Parameters-used-in}
\end{table}

\section*{Results: effectiveness of smartphone-based track-and-trace policies}

To evaluate differences in viral diffusion under alternative parameter combinations, for each simulated scenario we plot the overall proportion of the population infected at the peak and throughout the whole course of the epidemic. We use the baseline scenario of no tests and no app as point of reference to better understand the impact of the mitigations that are introduced. 

\subsection*{Testing without tracing}

\begin{figure}
	\includegraphics[scale=0.55]{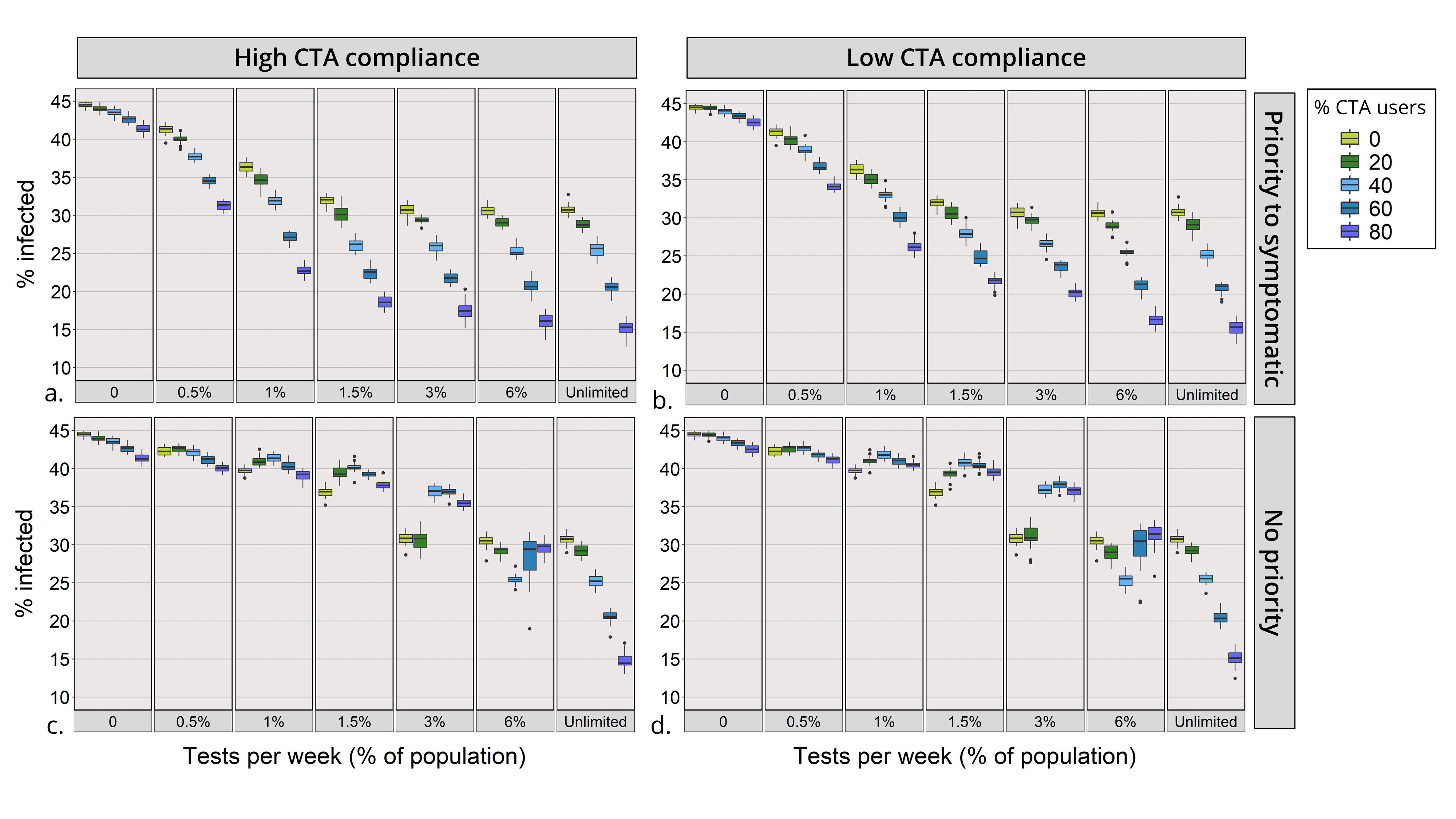}
	
	\caption{Percentage of the population infected
		during the course of epidemic. Scenarios vary by testing capacity
		(x-axis) and percentage of CTA users (boxplot\textquoteright s colour).
		Diagrams are organized by testing policy and compliance with self-isolation
		of CTA users: (a) High compliance and priority for testing symptomatic
		cases; (b) Low compliance and priority for testing symptomatic cases;
		(c) High compliance and no priority for testing symptomatic cases;
		(d) Low compliance and no priority for testing symptomatic cases.
		The boxplots show the median and interquartile range of multiple simulation
		runs.}
	\label{fig:Total-percentage-of}
\end{figure}

When testing is the only mitigation used (with no CTA), as testing capacity increases from 0 to 3\%, overall infections decreases from 44\% to 31\% (Figure \ref{fig:Total-percentage-of}) and infections at the peak are reduced by 59\% (Figure \ref{fig:Reduction-of-the}). A further increase in testing capacity above 3\% does not result in a further decrease in infections. This can be explained by the prevalence of infected agents per day, which reaches 3\% of the population at the peak of the epidemic (Figure \ref{fig:Infected-cases-and}a, light green line). Since around 50\% of the infected are symptomatic, a testing capacity of 3\% appears as sufficient to test all the symptomatic COVID-19 cases, other contacts who were notified (such as relatives and classmates) as well as the portion of influenza-like illness cases that seek testing. Thus, testing increases compliance with self-isolation which in turn reduces transmission. 

\subsection*{Introducing the CTA with a testing policy that prioritises symptomatic cases }

\begin{figure}
	\includegraphics[scale=0.7]{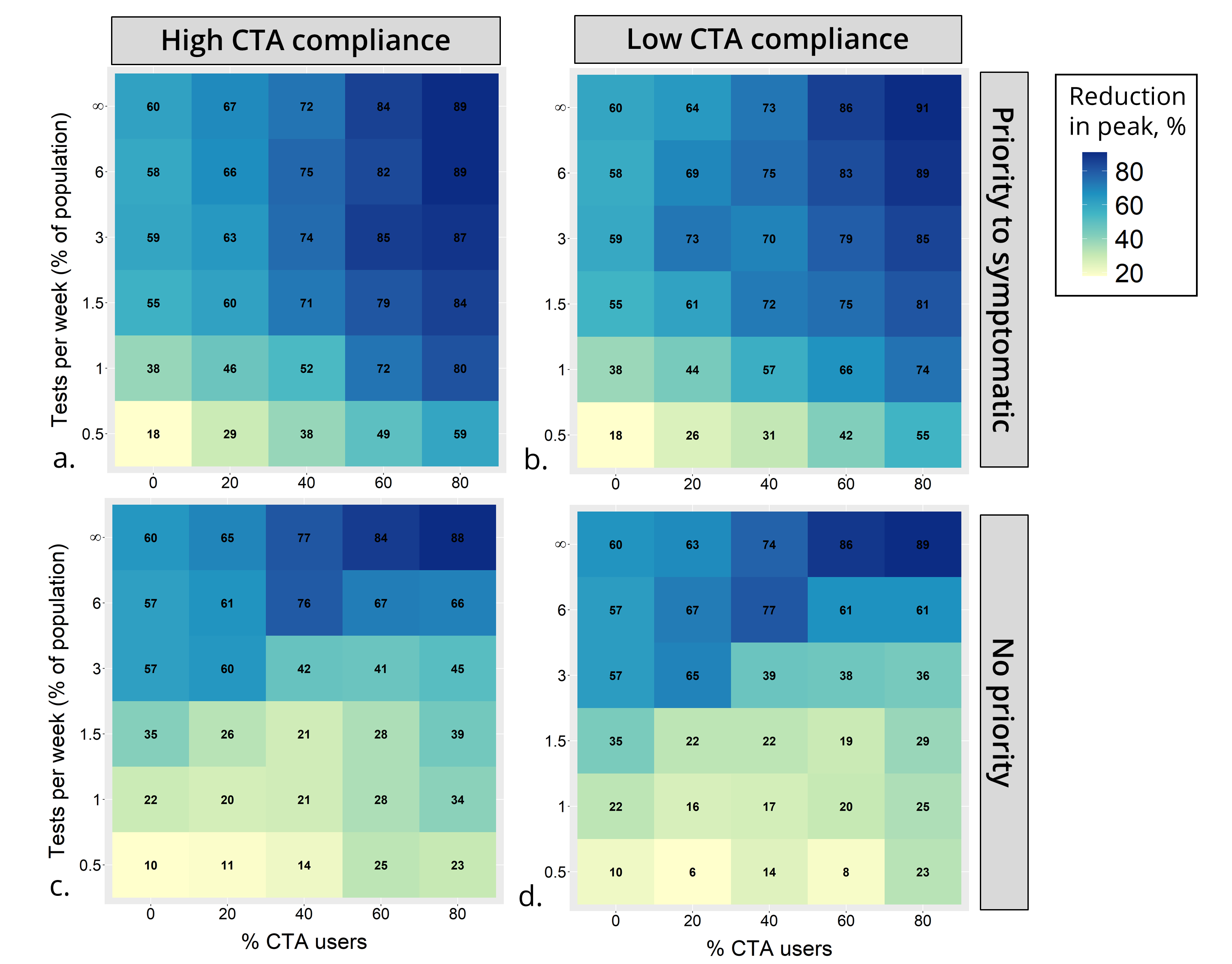}\caption{Reduction in infection prevalence at the peak
		of the epidemic. Reduction is relative to the peak of the epidemic
		in the baseline scenario and measured as percent reduction of this
		value. Scenarios vary by percent of CTA users (x-axis) and testing
		capacity (y-axis) and organised by testing policy and compliance with
		self-isolation of CTA users: (a) high compliance and priority for
		testing symptomatic cases; (b) low compliance and priority for testing
		symptomatic cases; (c) high compliance and no priority for testing
		symptomatic cases; (d) low compliance and no priority for testing
		symptomatic cases.}
	\label{fig:Reduction-of-the}
\end{figure}

Once the CTA is introduced, CTA users who are notified of having been in contact with an infected agent seek testing. When symptomatic agents are prioritised for testing, as the proportion of CTA users increases, overall infections throughout the epidemic decrease (Figure \ref{fig:Total-percentage-of}a-\ref{fig:Total-percentage-of}b), and so do infections at the peak of the epidemic (Figure \ref{fig:Reduction-of-the}a-\ref{fig:Reduction-of-the}b), irrespective of testing capacity. Note that some reduction in overall infections is observed as CTA adoption rates increase even in the scenario with no testing; in this case CTA notifications are emitted only by a fraction of app users who were severely ill and diagnosed in the hospital. Moreover, a synergy exists between testing and the CTA, resulting in a larger reduction in the spread of the virus, reflected in a decrease in infections both overall, and at the peak of the epidemic. 
The decrease is most substantial when testing is not limited. For example, when 80\% of the population uses the CTA, the percentage of the population infected throughout the epidemic decreases from 45\% to 15\% (Figure \ref{fig:Total-percentage-of}a-\ref{fig:Total-percentage-of}b) and cases at the peak of the epidemic reduce by 89\% (Figure \ref{fig:Reduction-of-the}a-\ref{fig:Reduction-of-the}b).
Moreover, for intermediate levels of CTA adoption (40\%-60\%) and testing capacity (1.5\%-3\%), overall infections decrease to 22\%-27\% of the population and cases at the peak reduce by 70\%-85\%, depending on the scenario (Figure \ref{fig:Total-percentage-of}a-\ref{fig:Total-percentage-of}b \& Figure \ref{fig:Reduction-of-the}a-\ref{fig:Reduction-of-the}b). 

\subsection*{CTA with no priority to test symptomatic cases }

\begin{figure}[h]
	\includegraphics[scale=0.55]{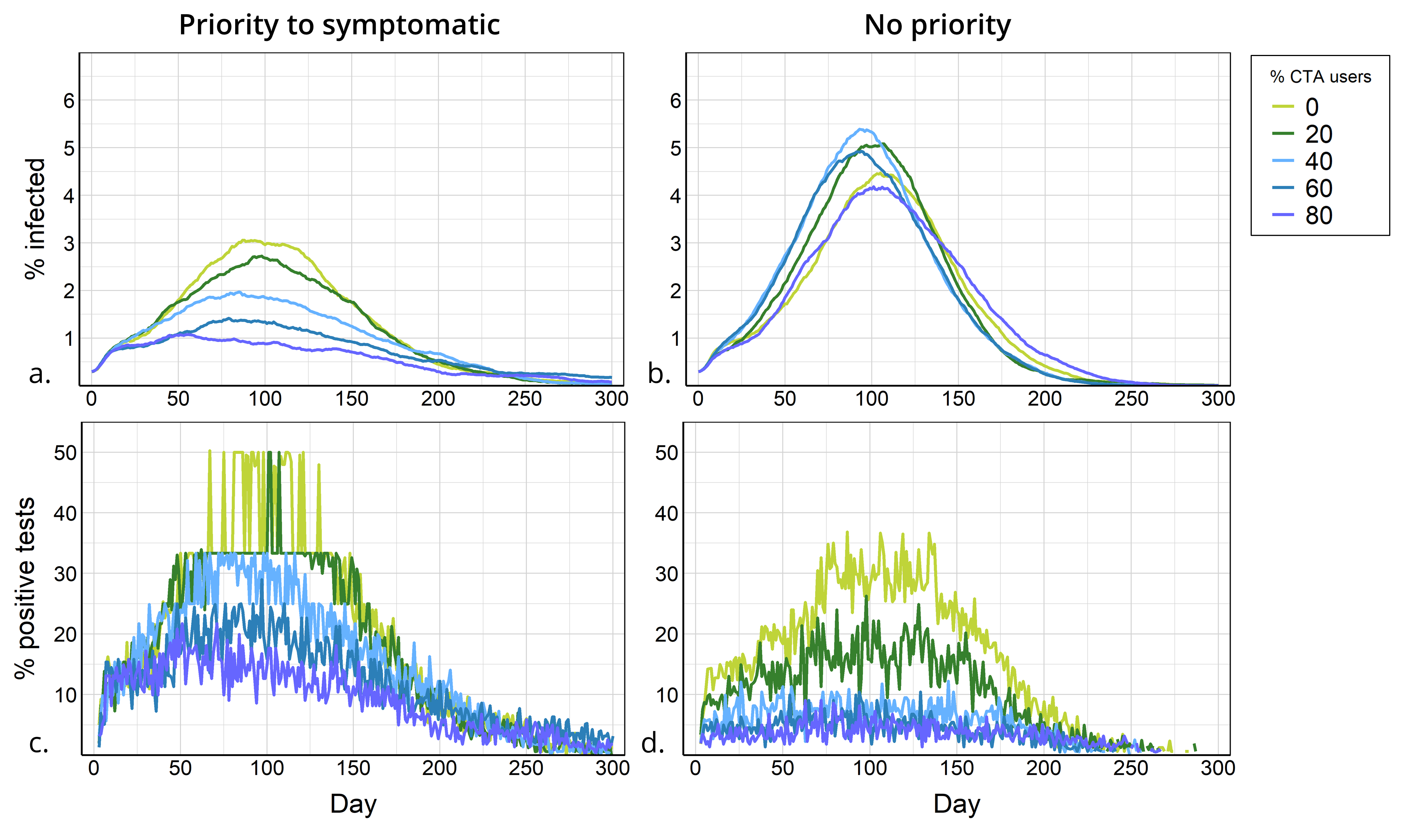}
	
	\caption{ Infection prevalence and positive tests for
		a testing policy with and without priority to symptomatic cases. Percentage
		infected (a) and percentage of positive tests (c) by day, for a testing
		policy with priority to symptomatic. Percentage infected (b) and percentage
		positive tests (d) by day, for a testing policy without priority.
		Percentage of CTA users in the scenario is marked by line colour.
		In these scenarios: testing capacity = 1.5\% and CTA compliance is high.
		Note that in the scenarios with no CTA users, although only symptomatic
		cases are tested, the percentage of positive tests is not 100\% because
		of cases with influenza-like illness that are also being tested. }
	\label{fig:Infected-cases-and}
\end{figure}

When the testing capacity is restricted and symptomatic agents are not prioritised for testing, an increase in the proportion of CTA users does not always lead to a decrease in infections (Figure \ref{fig:Total-percentage-of}c,
\ref{fig:Total-percentage-of}d). This is observed for all testing capacities between 0 and 6\%. In the case of a 3\% testing capacity, CTA adoption rates of 40\%
- 80\% result in substantially more infections than the scenario with no CTA users, both for high and low compliance with the CTA. This somewhat counterintuitive effect is explained by an inefficient testing policy. If symptomatic agents are not prioritised, they are as likely to be tested as those notified by the app, the majority of which are not infected. However, the increased demand for testing generated by CTA users who receive notifications depletes the stocks and prevents several infected cases from being detected. This phenomenon is apparent when comparing the \textquoteleft efficiency\textquoteright{} of the two testing policies (Figure \ref{fig:Infected-cases-and}c,
\ref{fig:Infected-cases-and}d). When symptomatics are prioritised the proportion of positive tests is higher for any given rate of CTA adoption, even though the prevalence of infection is lower than in the no priority scenario.

In general, the model suggests that the adoption of the CTA triggers two competing dynamics. On one hand, it informs agents and leads some of those infected to isolate even in the absence of a test: a crucial outcome, given the high proportion of a-symptomatic cases. On the other hand, it leads to a large and somewhat \textquoteleft inefficient\textquoteright{} depletion of tests, as several uninfected agents would likely seek testing after a notification from the app. The model shows that this counterproductive effect can be mitigated either with a testing policy that prioritises symptomatics cases, or with an overall increase in testing capacity. The scenarios with unlimited tests always show a substantial advantage in using the CTA.

\subsection*{CTA users\textquoteright{} compliance with self-isolation }

When compliance with self-isolation of CTA users is high, overall infections throughout the epidemic are only a few percentage points ($\sim$3\%-5\%) lower than when compliance level is low. The level of compliance with self-isolation becomes more influential the higher the proportion of CTA users, and at lower levels of testing capacity (Figure \ref{fig:Total-percentage-of}). E.g. in a scenario with 1\% testing and 80\% CTA users, overall infections reach 23\% when compliance is high and 27\% when compliance is low. The peak of the epidemic compared to the baseline scenario is reduced by \textasciitilde{} 80\% when compliance is high and \textasciitilde by 74\% when compliance is low. (Figure \ref{fig:Reduction-of-the}). 

\section*{Discussion}

The primary aim of this work is to contribute to the understanding of the complexity embedded in the interaction between the circulation of COVID-19 and the mitigations proposed. The results presented above show that app-based contact-tracing has the potential to mitigate the spread of COVID-19 in a social distancing scenario. With that,
CTA efficiency in reducing the spread relies on additional elements: testing capacity and management of limited testing resources. When priority for testing is given to symptomatic cases the impact of CTA is linear; the higher the levels of CTA adoption, the more the virus is suppressed. This is true for any level of CTA adoption and any
level of testing capability. 

A more complex dynamic emerges in the model under limited testing resources when symptomatic cases are not prioritised. Increases in the adoption of the CTA produce a spillover effect whereby the large number of (mostly non-infected) agents notified by the CTA depletes the testing stock and prevents a number of symptomatic agents from being tested. In this case, the CTA may even produce an increase in the infection rate, especially if testing capacity is not sufficiently high relative to the number of CTA users. 

These phenomena that emerge in the simulation can offer generalisable policy-relevant insights. First, the model shows that to optimise the contribution of the CTA towards epidemic control an adequate testing capacity has to be in place. This capacity is dependent on the proportion of CTA users in the population. Second, the specific way in which testing resources are managed substantially affects the effectiveness of containment. The suggestion is that governments should implement the CTA in parallel with a substantial increase in testing capacity, and accurately plan the details of the testing policy, keeping in mind that the mere availability of the CTA substantially alters patterns of demand for testing. 

One of the aims of the model was to explore the impact of certain behavioural factors: we have shown that, even under conditions of low compliance of CTA users towards self-isolation, a significant decrease in the spread of the virus is achieved. Moreover, compliance with self-isolation can be enhanced by increasing the availability
of testing to provide more certainty to exposed individuals regarding their infection state. Recent findings suggest that providing people with assurances about their livelihoods, by means of financial compensation, will increase compliance with self-isolation. \cite{Bodas2020}

As we demonstrated, in the more optimistic scenarios overall infections throughout the course of the epidemic are reduced to a third of the baseline scenario. The model shows that, when accompanied by social distancing,  CTA and testing are likely to reduce the spread of the epidemic and contribute significantly to lowering the epidemic peak. This is especially crucial for the functioning and manageability of the healthcare system until a vaccine and other pharmaceutical treatments are developed. Furthermore, the model suggests that the effectiveness of the CTA is higher in situations of higher viral circulation (Sensitivity analysis, Supplementary Information). This potentially increases its importance in controlling the epidemic at the post-lockdown stage where reopening of economic activity inevitably leads to higher infection rates.

\subsection*{Limitations}
The unprecedented nature of technology-based contact tracing, and the novelty of the disease itself, make it difficult to include, parametrise and validate every aspect of the model based on hard evidence, not least because such evidence often does not yet exist. As for any modelling exercise, the generated results rely on the model’s assumptions that simplify reality. We based the disease course parameters on emerging evidence as specified in Supplementary Table S1, while the individual level transmission parameters were calibrated to match the aggregated reproduction number ($R_0$) reported in the UK (Table \ref{tab:Contact-type-and}). Validation of the contact patterns emerging in the model was based on a comparison with a UK-based survey (Figure  \ref{fig:Epidemic-dynamics-with}). With that, even with proper calibration of the ABM to reported infection rates ($R_0$) it is possible that a different combination of individual based contact patterns and transmission rates occurring in different social environments could produce similar infection rates on the aggregated level \cite{axtell1994agent}. In such a case, the relative contribution of different social domains to the spread may differ and affect the extent to which CTA is employed. To address this issue, a sensitivity analysis for parameters affecting the viral spread is offered in the Supplementary Information section. The analysis shows that while the overall proportion of population infected may vary across parameter values, the findings regarding the impact of the CTA remain consistent for all scenarios. 
Another potential factor not implemented in our model is adoption rate differentials among demographic groups: the impact of such heterogeneity is still to be explored. A final caveat relates to our assumptions that the CTA functions “perfectly” (no technical malfunctions or notification delays) and that test results are accurate and returned within one day. In reality,  shortcomings in these matters may impair the contact tracing strategy. 

\section*{Conclusions}
To conclude, we maintain that smartphone-based contact-tracing is a viable epidemic mitigation strategy, worth pursuing on the part of governments. The model suggests that, as larger fractions of society adopt the CTA, the spread of the virus is increasingly reduced, and, therefore, the benefits extend to the wider population. In principle, the CTA offers speed and cost efficiencies that can complement and extend traditional manual contact-tracing methods. In our view, the idea of technology-based contact-tracing should not be dismissed, especially not on the grounds that it may widen inequality by penalising those with limited access to the technology, or fail protect the elderly population who is less likely to adopt it. Paradoxically, users of the CTA do not benefit directly from doing so, since it only operates when an individual may already have been exposed to the virus \cite{Rizzo2020}. While the CTA operates on a personal level by informing individuals of a possible risk of exposure, its general impact is at the societal level. By tracking exposed individuals and informing them to seek testing and self-isolate, transmission chains are interrupted, the spread of the virus is reduced, and so is the likelihood of infection for the whole population, including those who are not using the CTA.

\bibliography{covid.bib,nonpaper.bib}

\section*{Acknowledgements}

This work was supported by the Medical Research Council [grant numbers
MC\_UU\_12017/10 and MC\_UU\_12017/14] and Chief Scientist Office [grant number SPHSU10]. Both
grants provide 5 years of core research support for JA and SP.

\section*{Author contributions}

Both authors conceived and programmed the model, designed experiments, analysed the results, and reviewed the manuscript. 

\section*{Additional information}
 \subsection* {Competing interests} 
 The authors declare no competing interests. 

\end{document}


\title{Exploring the effectiveness of a COVID-19 contact tracing app using an agent-based model - Supplementary information}
\author{Jonatan Almagor, Stefano Picascia}
\maketitle

\section*{Disease state transition: probabilities, duration, literature sources}

\begin{table}[H]
\begin{tabular}{c>{\centering}p{0.2\columnwidth}>{\centering}p{0.45\columnwidth}c}
\toprule 
Parameter & Description & Value & Reference\tabularnewline
\midrule
\midrule 
$\alpha^{A}$ & Probability of infected agent of age A of becoming symptomatic & age group: 0-10, 10-20, 20-40, 40-50, 50-70, 70+

probability: 2\%, 26\%, 55\%, 62\%, 72\%, 82\% & \cite{Davies2020}\tabularnewline
\midrule 
$\delta^{A}$ & Probability for symptomatic agent of age A to progress to severe disease & age group: 0-15, 15-40, 40-50, 50-60, 60-70, 70+

probability: 2\%, 6\%, 9\%, 13\%, 17\%, 20\% & \cite{Verity2020}\tabularnewline
\midrule 
$\gamma^{A,G}$ & Probability of death for severely ill agent of age A and gender G & age group: 0-15, 15-40, 40-50, 50-60, 60-70, 70+

male probability: 0.5\%, 3\%, 8\%, 9\%, 16\%, 25\%, 50\%

female probabilities by age corresponds and reduced by 20\% & \cite{PHE2020}\tabularnewline
\midrule 
$d_{inc}$ & Incubation period & Value drawn from Gamma distribution (5.1, 1) & \cite{Lauer2020}\tabularnewline
\midrule 
$d_{asy}^{A}$, $d_{mild}^{A}$ & Disease duration of asymptomatic and mild symptomatic agent of age
A & age group: 0-40, 40-50, 50-60, 70+

mean duration, days: 8, 12, 15, 20. Value drawn from normal distribution
with the age group mean and $SD=0.25*mean$ & \cite{Chen2020}\tabularnewline
\midrule 
$d_{sev}$ & Duration of severe disease before hospital admission & Gamma distribution (6.5, 0.9) mean = 7 days & \cite{Perez-Guzman2020}\tabularnewline
\midrule 
$d_{hos}^{A}$ & Length of hospital stay & age group: 0-40, 40-50, 50-60, 70+

mean duration, days: 8, 12, 15, 20

Value drawn from a normal distribution with the age group mean and

$SD=0.25*mean$ & \cite{Perez-Guzman2020}\tabularnewline
\bottomrule
\end{tabular}\caption{Disease state transition probabilities and duration, with literature reference}
\label{tab:Disease-state-transition}
\end{table}
\section*{Sensitivity analysis}
In order to test whether the impact of the CTA on transmission dynamics as demonstrated in our simulation still holds under various assumptions of contact patterns in the population and under different transmission probabilities, we conducted a sensitivity analysis to selected parameters as specified in Table \ref{sensitivity}. The sensitivity analysis was carried out assuming a social distancing scenario and a testing policy that prioritises symptomatic agents when testing capacity is 1.5\% of the population per week, and unlimited.   
The results are presented in the figures below. In all the figures, the black trajectory represents the value used in the simulations discussed in the main paper. 

\begin{table}[h]
\begin{tabular}{|c|>{\centering}p{0.2\columnwidth}|>{\centering}p{0.2\columnwidth}|>{\centering}p{0.2\columnwidth}|c|}
\hline 
Parameter & Description & Value in model & Range tested for sensitivity & Figure\tabularnewline
\hline 
\hline 
$\beta_{c}$ & Transmission probability per network contact & 0.056 & 0.028 -- 0.084 & Fig. \ref{fig:Sens-beta-c} \tabularnewline
\hline 
$p$ & Percentage of local area population that agents meet in random encounters & 0.7\% & 0.35\% -- 1.4\% & Fig. \ref{fig:Sens-prop-random}  \tabularnewline
\hline 
$\beta_{r}$ & Transmission probability per random contact  & 0.0056  & 0.0028 -- 0.0112  & Fig. \ref{fig:Sens-betaR} \tabularnewline
\hline 
$f$ & Number of friends that agents meet per encounter & Random draw from 1-10\% of agent's ties & Varying no. of friends by:
-50\% -- +100\% & Fig.  \ref{fig:Sens-friends} \tabularnewline
\hline 
\end{tabular}

\caption{Parameters and range of values tested in sensitivity analysis}

\label{sensitivity}
\end{table}

The diagrams show that, as expected, varying each of the parameters does influence the overall number of infections, however the main model outcome - that higher CTA adoption rates always translate in lower infections - is consistently emerging. The influence of the CTA can be observed by the negative slope in all the figures. The steepness of the slope depends on the degree to which CTA adoption decreases the spread.

In particular, the sensitivity analysis shows that the CTA is more effective in scenarios of higher viral circulation (more contacts or higher infectiousness). At the higher values of all the parameters tested more infections are generated than those in the base model. For these conditions a steeper slope is observed,  indicating a higher relative reduction in infections as CTA adoption rates increases. This dynamics can be be explained: in higher transmission conditions each infectious agent infects more susceptible agents (on average) compared to conditions of low transmission; therefore each infected case who is also a CTA user sends alerts to more agents who were exposed to him and were infected, which in turn reduces their exposure to the population (by self-isolating); this translates into a relatively higher reduction in infections. On the contrary, for low conditions of transmission the viral circulation is reduced overall, and therefore the effect of the CTA is less apparent, because it is activated in less infection occasions.

This effect is particular noteworthy as countries around the world enter the second wave of the pandemic and at the same time are trying to avoid strict lockdown and maintain sections of the economy open, which results in ongoing interactions between people. Under these circumstances high CTA adoption rates in the population combined with sufficient testing capacity can significantly reduce the spread of COVID-19.   

\begin{figure}[H]
    \centering
    \includegraphics[scale=0.4]{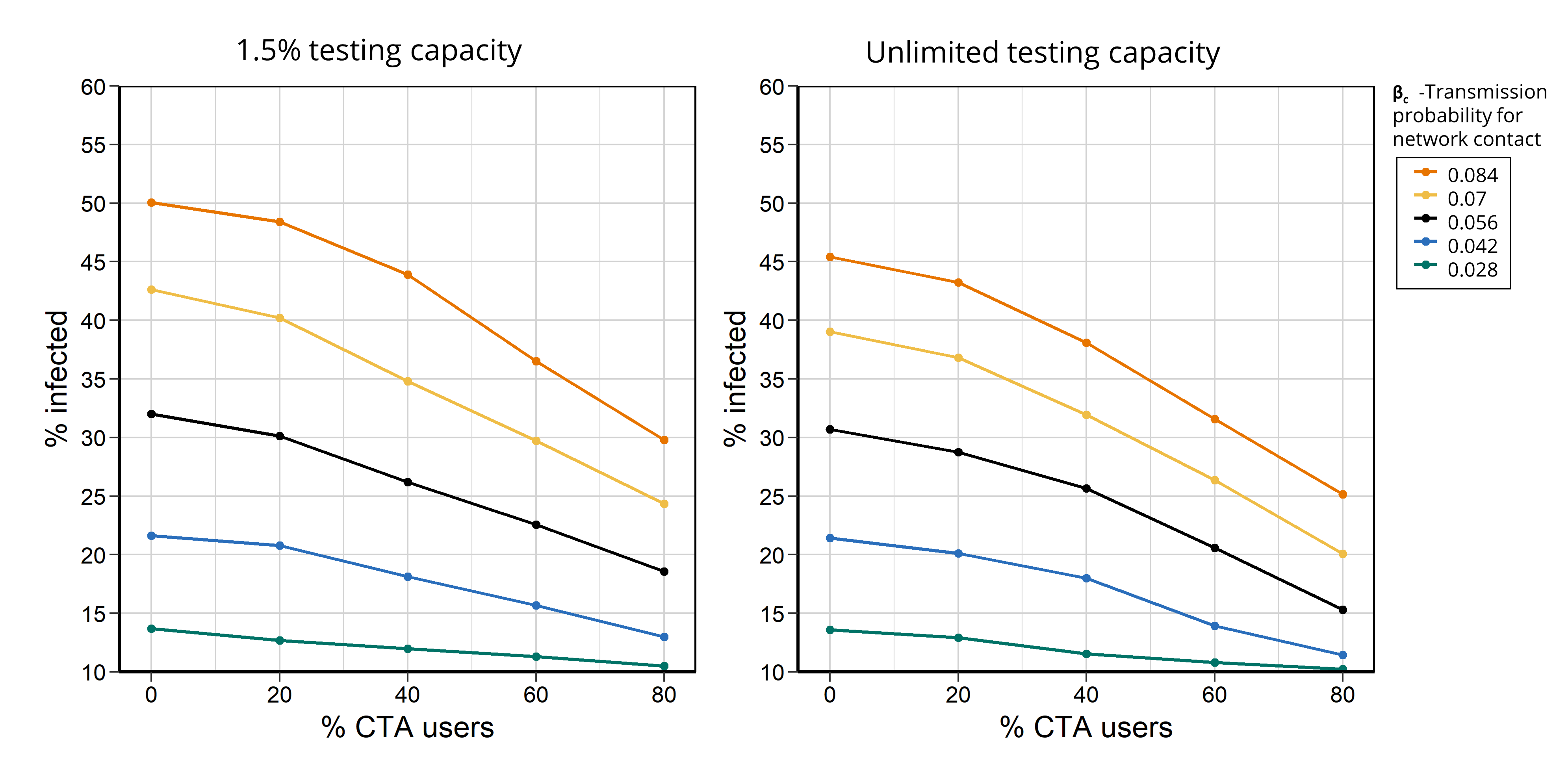}
    \caption{Sensitivity analysis for transmission probability per network contact (friends and work colleagues) ($\beta_{c}$). For each $\beta_{c}$ value used in the simulation we present the percentage of the population infected during the course of the epidemic (y-axis) for varying rates of CTA users (x-axis). Values of $\beta_{c}$ are presented by lines with unique colours. Black line represents the value used in paper. The slope of the trajectory represent the influence of the CTA; a steeper negative slope represent a larger relative reduction in infections. Scenarios with testing capacity of 1.5\% (left plot) and unlimited testing capacity (right plot).}
    \label{fig:Sens-beta-c}
\end{figure}

\begin{figure}[H]
    \centering
    \includegraphics[scale=0.4]{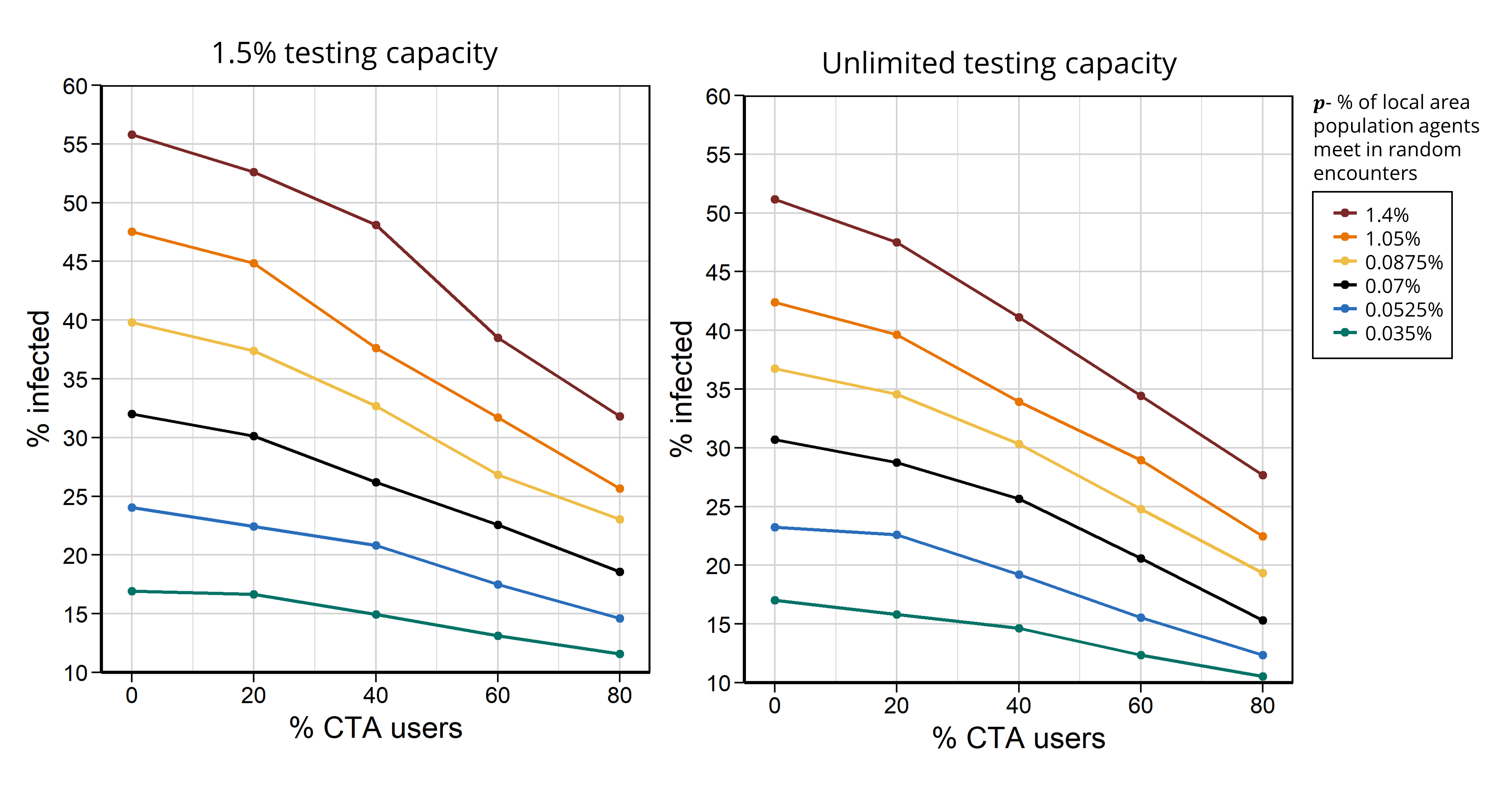}
    \caption{Sensitivity analysis for percentage of local area population that agents meet in random encounters ($p$). For each $p$ value used in the simulation we present the percentage of the population infected during the course of the epidemic (y-axis) for varying rates of CTA users (x-axis). Values of $p$ are presented by lines with unique colours. Black line represents the value used in the paper. The slope of the trajectory represent the influence of the CTA; a steeper negative slope represent a larger relative reduction in infections. Scenarios with testing capacity of 1.5\% (left plot) and unlimited testing capacity (right plot).}
    \label{fig:Sens-prop-random}
\end{figure}
\begin{figure}[H]
	\centering
	\includegraphics[scale=0.4]{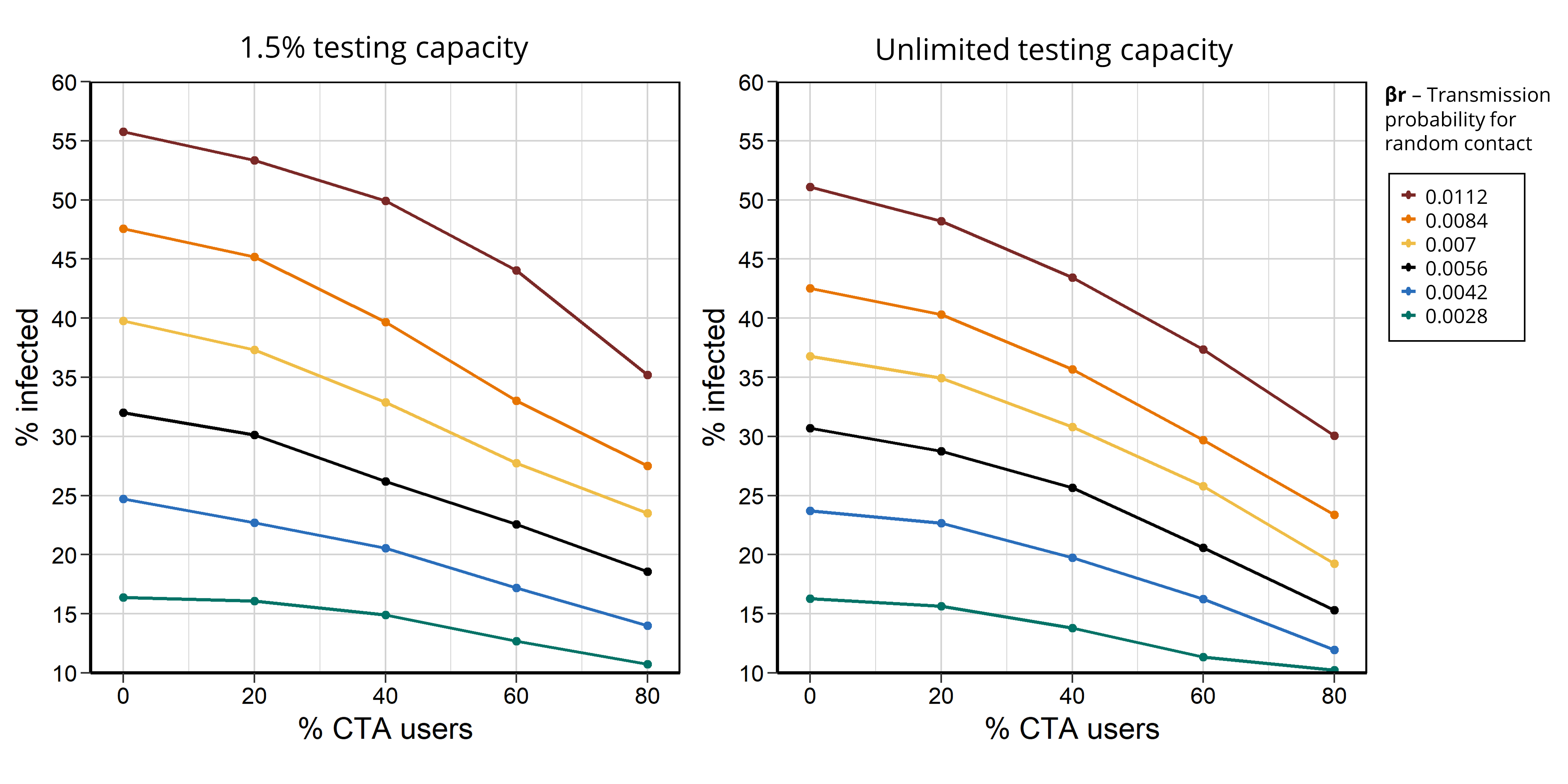}
	\caption{Sensitivity analysis for transmission probability per random contact ($\beta_{r}$). For each $\beta_{r}$ value used in the simulation we present the percentage of the population infected during the course of the epidemic (y-axis) for varying rates of CTA users (x-axis). Values of $\beta_{r}$ are presented by lines with unique colours. Black line represents the value used in paper. The slope of the trajectory represent the influence of the CTA; a steeper negative slope represent a larger relative reduction in infections. Scenarios with testing capacity of 1.5\% (left plot) and unlimited testing capacity (right plot).}
	\label{fig:Sens-betaR}
\end{figure}
\begin{figure}[H]
    \centering
    \includegraphics[scale=0.4]{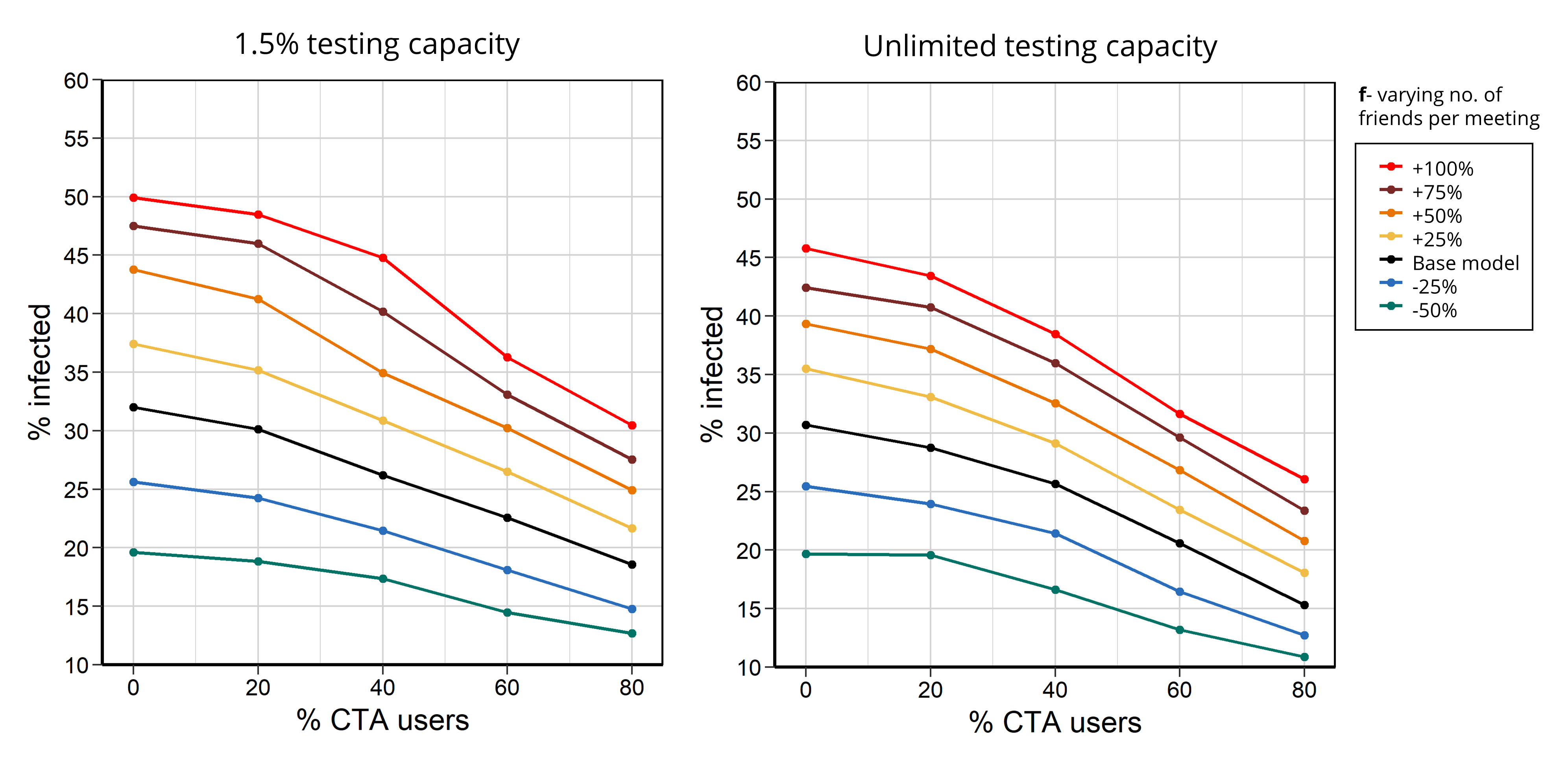}
    \caption{Sensitivity analysis for number of friends per social encounter ($f$). For each $f$ value used in the simulation we present the percentage of the population infected during the course of the epidemic (y-axis) for varying rates of CTA users (x-axis). Values of $f$ are presented by lines with unique colours. Black line represents the value used in the paper. The slope of the trajectory represent the influence of the CTA; a steeper negative slope represent a larger relative reduction in infections. Scenarios with testing capacity of 1.5\% (left plot) and unlimited testing capacity (right plot).}
    \label{fig:Sens-friends}
\end{figure}

\section*{Pseudo code of model step-function}

The full source code and supporting datasets of the model are available
at the following address: \url{https://github.com/harrykipper/covid} For ease of interpretation we offer a simplified version of the step function and the main infection
process in the form of pseudocode, listed below.

\begin{lstlisting}

function step(): 
tests-available = tests-available + tests-per-day
for (i in contacts_stored_in_CTAs) {
	days(i) = days(i) + 1
	if days(i) > 10 {remove_contact(i)}
	}

for (i in all_agents){
	if infected(i) {
		diseaseProgress()
		if asymptomatic(i) {b = b * Decay}
		if symptomsAppear {seek-testing} 
		infect()
	}
	if crowd_worker(i){meet_customers}
}

function infect(): 
if isolating {b_hh = b_hh * 0.7}
for (i in household_members){ 	
	if b_hh > random-float(1) {infect i}
	}

if not isolating {
	for (k in poisson (p * no_zone_residents)) {
		if cta_user(self) and cta_user(k) 
			{new_app_contact(self,k)}
		if (b_r * age_discount) > random-float(1)
			{infect k}
	}
    if meeting-relatives{
       if b_c >random_float(1) {infect random-relative }
     }
	if office_worker {
		for (j in c_colleagues) {
			if cta_user(self) and cta_user(j) 
				{new_app_contact(self,j)}
			if b_c > random-float(1) {infect j}
		}
		if cta_user(self) and cta_user(random_other_colleague)
			{new_app_contact(self,random_other_colleague)}
		if b_c > random-float(1) {infect random_other_colleague}
	} else if in_school {
		for (c in 1 to (no_of_classmates/2)){
			if b_c * age_discount > random-float(1) 
				{infect c}
			}
		}
	for (f in random(1,(no_of_friends * 0.1))){
		if cta_user(self) and cta_user(f)
			{new_app_contact(self,f)}
		if b_c * age_discount > random-float(1) {infect f}
	}
}

function meet_customers():
for (k in poisson (3p * no_of_zone_residents)) {
	if cta_user(self) and cta_user(k){ 
		new_app_contact(self,k)
	}
	if infected() {
		if (b_r * age_discount) > random-float(1) 
			{infect k}
	} else {if infected(k) {
		if (b_r * age_discount) > random-float(1) 
			{infect self}
	}
}

function seek_testing():
if tests-available > 0 { get-tested() } 
	else { decide_whether_to_isolate() }

function get-tested(): 
tests-available = tests-available - 1 
if positive {
	for (i in household_members) {i.decide_whether_to_isolate()} 
	for (i in relatives) {i.decide_whether_to_isolate()}
	if in_school {
		for (i in classmates) {i.isolate()}}
	if hasApp {
		for (i in app_contacts) {i.seek_testing()}
	}
}
\end{lstlisting}

\bibliographystyle{pnas2009}
\bibliography{covid,nonpaper}